\RequirePackage{etex} 
\documentclass[conference]{IEEEtran}
\IEEEoverridecommandlockouts
\usepackage{cite}
\usepackage{amsmath,amssymb,amsfonts}
\usepackage{mathtools}
\usepackage{bbm}
\usepackage{pifont}
\usepackage{textcomp}
\usepackage{xcolor}
\usepackage{physics}
\usepackage{lipsum}
\usepackage{ifthen}
\usepackage[colorlinks=true,urlcolor=teal,
            linkcolor=teal,citecolor=teal]{hyperref}
\usepackage{standalone}
\usepackage{tikz}
\usepackage{graphicx}
\usepackage{yquant}
\usepackage[utf8]{inputenc}
\usepackage{pgfplots}
\usepackage{pgfmath}
\DeclareUnicodeCharacter{2212}{−}
\usepgfplotslibrary{groupplots,dateplot}
\usetikzlibrary{patterns,shapes.arrows,math}
\usetikzlibrary{positioning}
\pgfplotsset{compat=newest}
\usepackage{csquotes}
\usepackage{caption}
\usepackage{subcaption}
\usepackage{tikz-network}
\usepackage{xspace}
\usepackage{multirow}
\usepackage{booktabs}
\usepackage{bm}

\usepackage{fancyhdr}

\newcommand{\eg}{\emph{e.g.}\xspace}
\newcommand{\ie}{\emph{i.e.}\xspace}
\newcommand{\etal}{\emph{et al.}\xspace}

\newcommand{\ibmq}{IBM-Q\xspace}
\newcommand{\ionq}{IonQ\xspace}
\newcommand{\rigetti}{Rigetti\xspace}

\newcommand{\swpgates}{\textsc{Swap} gates\xspace}
\newcommand{\Cgate}[1]{\ensuremath{\text{C\raisebox{0.08em}{--}}\!#1}\xspace}
\newcommand{\suppweb}{\href{https://github.com/QSW2023NoiseModeling/QSW2023Noise_Modeling}{supplementary website}\xspace}
\renewcommand{\rho}{\varrho}

\definecolor{lfd1}{HTML}{FFFFFF} 
\definecolor{lfd2}{HTML}{E69F00}
\definecolor{lfd3}{HTML}{999999}
\definecolor{lfd4}{HTML}{009371}
\definecolor{lfd5}{HTML}{beaed4}
\definecolor{lfd6}{HTML}{ed665a}
\definecolor{lfd7}{HTML}{1f78b4}

\newboolean{anonymous}
\setboolean{anonymous}{false}

\begin{document}
\bstctlcite{IEEEexample:BSTcontrol}
\fancyhf{}
\renewcommand{\headrulewidth}{0pt}
\renewcommand{\footrulewidth}{0.5pt}
\fancyfoot[C]{}
\fancypagestyle{FirstPage}{
	\lfoot{\fontsize{8}{9} \selectfont\copyright2023 IEEE. Personal use of this material is permitted. Permission from IEEE must be
obtained for all other uses, in any current or future media, including
reprinting/republishing this material for advertising or promotional purposes, creating new
collective works, for resale or redistribution to servers or lists, or reuse of any copyrighted
component of this work in other works.}
}

\title{Effects of Imperfections on Quantum Algorithms:\\
A Software Engineering Perspective}

\author{
\ifbool{anonymous}{Anonymous author(s)}{
\IEEEauthorblockN{Felix Greiwe}
    \IEEEauthorblockA{\textit{Technical University of} \\
      \textit{Applied Sciences Regensburg}\\
      Regensburg, Germany \\
      \href{mailto:felix.greiwe@othr.de}{felix.greiwe@oth-regensburg.de}}
\and
  \IEEEauthorblockN{Tom Krüger}
  \IEEEauthorblockA{\textit{Technical University of} \\
    \textit{Applied Sciences Regensburg}\\
    Regensburg, Germany \\
    \href{mailto:tom.krueger@othr.de}{tom.krueger@oth-regensburg.de}}
\and
  \IEEEauthorblockN{Wolfgang Mauerer}
  \IEEEauthorblockA{\textit{Technical University of}\\
    \textit{Applied Sciences Regensburg} \\
    \textit{Siemens AG, Technology}\\
    Regensburg/Munich, Germany \\
    \href{mailto:wolfgang.mauerer@othr.de}{wolfgang.mauerer@othr.de}}
}
}

\maketitle

\begin{abstract}
Quantum computers promise considerable speedups over classical approaches, which has
raised interest from many disciplines. Since any currently available implementations
suffer from noise and imperfections, achieving concrete speedups for meaningful
problem sizes remains a major challenge. Yet, imperfections and noise may remain
present in quantum computing for a long while. Such limitations play no role in 
classical software computing, and software engineers are typically not well accustomed
to considering such imperfections, albeit they substantially influence core properties
of software and systems.

 In this paper, we show how to model imperfections with an approach tailored 
 to (quantum) software engineers. We intuitively illustrate, using numerical simulations,
 how imperfections influence core properties of quantum algorithms on NISQ systems,
 and show possible options for tailoring future NISQ machines to
 improve system performance in a co-design approach.
 Our results are obtained from a software framework that we provide in form
 of an easy-to-use reproduction package. It does not require computer scientists
 to acquire deep physical knowledge on noise, yet provide tangible and intuitively
 accessible means of interpreting the influence of noise on common software
 quality and performance indicators.
\end{abstract}

\section{Introduction}\label{sec:Introduction}
\thispagestyle{FirstPage}Quantum computing promises improvements and computational speedups over classical approaches for 
many tasks and problems, which include cryptography~\cite{ekert1992quantum}, machine 
learning~\cite{biamonte2017quantum}, optimisation~\cite{farhi2014quantum,bayerstadler:21}, or simulating 
chemical and physical systems~\cite{PRXQuantum.2.017003}. This has raised considerable interest
across scientific communities, including (quantum) software engineering---programmable quantum
computers and appliances will, eventually, involve software in one form or another.

Given the current state of available noisy intermediate scale quantum (NISQ) 
hardware~\cite{Preskill_2018}, actual quantum advantages are rarely seen (except for specially 
crafted problems~\cite{Arute:2019,Zhong:2020}). While error correction techniques for quantum computers 
exist, the required hardware resources exceed current system dimensions by many orders of 
magnitude~\cite{Roffe:2019}. Therefore, imperfections in quantum computers will be present in the 
forseeable future, and it is important for SW engineers and researchers to be aware how low
level effects like noise influence software qualities. 
Given these conditions, evaluating, characterising and predicting functional and non-functional
properties of quantum software is a complex, multi-facetted endeavour that
requires catering for many details, many of them are unaccustomed from
classical software engineering. It is becoming increasingly clear that possible performance 
benefits of quantum systems will be available only under particular circumstance that concern 
both, algorithms and hardware. Seemingly straightforward approaches (or naive analogies with 
classical systems and software) can quickly lead to bogus, unreliable or downright wrong 
statements that mis-characterise potential benefits of quantum approaches.
Our paper is intended to help software engineers develop a realistic expectation regarding the performance
of quantum algorithms under noise, on different types of hardware. We provide illustrative examples that
show impacts on a number of seminal (classes of) algorithms---Grover  search, quantum Fourier transform, 
and variational quantum circuits. Our main contributions are as follows:

\begin{itemize}
    \item We provide a self-contained exposition on modelling noise and imperfections 
    tailored at computer scientists and software engineers to create a
    tangible bridge between fundamental physics and non-functional
    properties conventionally employed in software engineering.
    \item Using numerical simulations, we show the influence of typical
    imperfections on multiple seminal algorithms for different hardware classes,
    and provide an intuitive understanding on potentials of hardware-software 
    co-design for future quantum computing systems.
    \item We provide a reproduction package~\cite{mauerer:22:q-saner} on the \suppweb (link in
	    PDF)\footnote{DOI-compliant version: \url{https://doi.org/10.5281/zenodo.8001512}} 
    that allows software engineers to quickly evaluate how noise and imperfections 
    influence their designs, without having to acquire a deeper understanding of low-level physical details.
\end{itemize} 

Our contribution intends to increase awareness in the quantum
software community on the impact of noise and imperfections on algorithmic performance, but also on the opportunities of co-designing future (NISQ) systems
whose properties are favourable for specific classes of applications.
In contrast to existing work, our paper places stronger focus on providing self-contained
instructions on \emph{how} to understand and model imperfections, and how to judge their
influence on key qualities or requirements of software and software architectures.
This is, for instance, required to support a well-informed discussion on finding proper
levels of abstraction needed to decouple peculiarities of QPUs as good as possible, yet
should not stand in the way of utilising the computational power provided by
QPUs. Likewise, knowledge of imperfections at a reasonable level
of detail can help researchers to avoid placing inflated expectations on the 
capabilities of quantum approaches.

The code in the reproduction package is based on the open source framework 
Qiskit~\cite{Qiskit}, and does not depend
on any real quantum hardware or proprietary compilers, which makes it accessible to a wide audience.
It not only enables researchers to easily re-create our results, but has,
instead, especially been designed to enable researchers and software engineers 
to extend it with own algorithms (and test/benchmarking cases), and study them
under the influence of various types of noise and gate sets, without having to manually implement the required physics-centric evaluation mechanisms.

The remainder of this paper is structured as follows: After reviewing related work
in quantum software engineering and quantum noise in Sec.~\ref{sec:related},
we discuss important characteristics of current QPUs, as well as
particularly relevant open hardware challenges, in Sec.~\ref{sec:QC_hardware}.
We provide a gentle introduction to modelling noise and imperfections
tailored towards software engineers in Section~\ref{sec:modelling},
and illustrate these considerations by discussing their impact on
several seminal quantum algorithms in Sec.~\ref{sec:simulations}, followed
by a discussion of the implications for software engineering in Sec.~\ref{sec:implications}. We conclude in Sec.~\ref{sec:discussion}.

\section{Related Work}\label{sec:related}
Since quantum software engineering is in its initial stages (yet, 
Piattini~\etal~\cite{piattini2021quantum} go as far as to proclaim 
a new \enquote{golden age} of software engineering), the available literature
still is sparse, and noise and imperfections are ignored in (or deemed irrelevant
for) many expositions that concentrate on possible future higher-level abstractions to 
quantum software engineering and quantum programming. For instance,
Perez-Castillo~\etal~\cite{Perez-Castillo:2021} 
discuss how to extend the unified modeling language to quantum circuits. 
Similarly, Gemeinhardt~\etal~\cite{GemeinhardtModelDrivenQ} suggest model-driven quantum software 
engineering as an  abstraction that extends established SWE methods.
Differences between quantum and classical engineering in terms of bug patterns
are studied by~Campos and Souto~\cite{Campos:2021}, as well as Zhao~\cite{Zhao:2021}.
Zhao~\cite{Zhao:2020} provides a detailed review of the available literature.
Piattini~\etal~\cite{piattini2020talavera} suggest principles for the future 
development of quantum software engineering, and highlight hybrid algorithms and the 
desirable independence of specific quantum software frameworks. 
Leymann~\etal~\cite{leymann2020bitter} focus on often ignored aspects of imperfections in quantum computing. 
Structured approaches for benchmarking software on quantum computers are considered
by Becker~\etal~\cite{Becker:2022} and Tomesh~\etal~\cite{Tomesh:2022};
in particular, the approach by Resch~\etal~\cite{Resch:2022} especially highlights the
importance of choosing appropriate noise models. Salas~\etal~\cite{salas2008noise}
consider noise effects on Grover's algorithm and state error thresholds. 

The performance of NISQ-era variational quantum algorithms,
particularly in the QAOA family, has been subject to intensive research; recent results 
include Refs.~\cite{Alam:2020,Wang:2021}. Other application fields like 
machine learning (see, \eg, Refs.\cite{Hubregtsen:2022, franz:22:qrl}) have
received similar consideration from an algorithmic benchmarking and performance
analysis point of view. Interestingly, it is known that noise need not necessarily
be detrimental, but can also contribute improvements, as recent research (\eg, 
Refs.~\cite{Liu:2022,guimaraes:2023}) demonstrates.
The physics-centric literature on quantum noise is
extensive, and reaches considerably further back; the seminal
exposition by Gardiner and Zoller~\cite{Gardiner:2000} contains many of the fundamental results.
Noise and imperfections in all possible implementation platforms for quantum
computers from a physical point of view have likewise been considered in substantial depth and breadth,
for which Bharti~\etal~\cite{Bharti:2022} provide a review.

Characterising the capabilities of quantum computers is, in general, an active
field of research: Considerations based on cross entropy~\cite{Boixo:2018} and quantum volume~\cite{Cross:2019} consider properties of random circuits, and aim at
a generically usable comparison metric that is applicable across implementation 
techniques, but does not allow for deriving concrete statements on algorithms
or use-case scenarios. Application oriented benchmarks (\eg, Refs.~\cite{Lubinski:2021,Lubinski:2023,krueger:20:icse}), and other domain-specific (\eg,~\cite{Alexander:2022,schoenberger:23:leap}) or generic (\eg, Refs.~\cite{Li:2022,Koen:2022,Finzgar:2022}) approaches, consider more
concrete perspectives, but often use techniques that are unaccustomed for
software engineers. We aim, in contrast, at a correct, yet tangible and
algorithm-oriented approach that is accessible and useful for the
software engineering community.

\section{Quantum Hardware and HW Challenges}\label{sec:QC_hardware}
One major challenge in quantum computing is to provide an isolation between
the fragile quantum bits that carry quantum information, and are used to
perform computations on, and the surrounding environment. Interactions
between qubits and the environment lead to the loss of quantum information
(\emph{decoherence}), and therefore degrade the quality of computational
processes. Likewise, operations on one or more of the qubits that perform
the actual computation may be imperfect, and usually implement a transformation 
that is only \enquote{close to} the actual specification, including random variations.
Both aspects do not occur in  classical systems (or can be very well countered), and 
correspondingly, software engineers (outside, probably, highly specialised domains like 
safety-critical engineering) need not be concerned with the corresponding phenomena.

It is still unclear which basic physical concepts will provide the basis of
future quantum computers. A multitude of possible approaches are currently 
developed and investigated, including systems based on trapped ions, neutral atoms,
superconducting semiconductor-based implementations, or
photonic systems.

We chose two common architectures of commercial interest to highlight the essential,
far-reaching differences in their physical implementation that, as we will argue in
this paper based on numerical simulations, substantially impact many properties of
systems that are relevant to software engineering.

\subsection{Physical Foundations}
\subsubsection{Trapped Ions}
By using an electromagnetic field to hold ions together in a trap, they
serve as building blocks to realise qubits by using stable (internal) 
electronic states of the ions together with so-called (external) collective 
quantised motion states of all ions assembled in the trap.
Laser pulses are used to control and couple the internal and motion states,
realise single- and multi-qubit gates, and cool (slow down) the ions to
motional lowest-energy states (see. \eg, 
Ref.~\cite{bruzewicz2019trapped} for details).

A salient characteristic of the motional coupling that affects all
quantum bits is that two-qubit operations can be executed between any
two qubits, which means that the system provides a fully-meshed
coupling graph. Likewise, very high gate fidelities and coherence
times can be achieved, as summarised in \autoref{tab:hwdata}.
Despite laser cooling of the involved ions, the overall system operates
at room temperature~\cite{Pogorelov:2021}.

In contrast to these advantages, gate execution times are 
comparatively large; operations require microseconds of processing
time. Additionally, it is not straightforward to scale trapped ion systems 
to more than, say, 100 qubits, while maintaining the motional coupling
between ions. Noise in gate application arises from variations of intensities
and phases of of the lasers involved, but also from external electromagnetic
fields that cannot be completely shielded off.

\subsubsection{Superconducting Transmons}
Superconducting quantum computing exploits quantum mechanical properties of macroscopic
structures that stem from Cooper electron pairs that form at very low temperatures in
superconductors.

The need for such low temperatures is a disadvantage compared to trapped ion systems.
Also qubits are coupled to qubits in their direct neighbourhood, and achievable coherence times are 
orders of magnitudes lower. However, gate times in superconducting devices are in the
nanosecond range, and larger systems (in terms of qubits count) can be built compared to
trapped ion system. Finally, the manufacturing process can benefit from established 
industrial semiconductor know-how.

\subsection{Challenges}
Several limitations of current quantum computers extend across physical realisations
and need to be considered when designing quantum software components, or when
planning experiments to judge feasibility or scalability of proposed quantum architectures.
Some of the limitations are specific to NISQ systems, others
concern intrinsic limitations of quantum computers and algorithms that need to be
taken into account for any consideration relating to quantum software engineering
or architecture. Noise (\ie, effects of imperfect quantum information representation
and manipulation), limited connectivity between qubits, and gate timing characteristics
offer substantial potential for future engineering improvements; they can be seen
as hardware \enquote{parameters} to a certain extent from a SWE point of view. We study the
respective potentials in detail the next section.

\subsubsection{Noise}
\label{sec:hw_noise}
Quantum states are fragile, and operations on such states require involved physical 
manipulation techniques that are hard to implement perfectly---any real-world
implementation slightly deviates from a theoretically desired perfect operation. Likewise,
information in quantum states is perturbed by interaction with an outside environment,
which is unavoidable because of the need to interact with and manipulate the states
to perform computations. The (in)stability of quantum states and quantum operations
is characterised by established measures that we discuss below; representative
measures for three different commercially accessible platforms are shown in 
Table~\ref{tab:hwdata}.

The \emph{coherence times} $T_1$ and $T_2$ indicate how resilient the information stored
in qubits is against perturbations (longer times are better). \(T_{1}\) gives the average time 
it takes a qubit to \enquote{relax} from $\ket{1}$ to state $\ket{0}$ (bit flip).

The stability of the relative phase in a superposition state \(\ket{+} = 1/\sqrt{2}
(\ket{0}+\ket{1})\) is quantified by \(T_{2}\), providing the average time after which \(\ket{+}\) has evolved into an equal-probability classical mixture of \(\ket{+}\) and \(\ket{-} = 1/\sqrt{2}(\ket{0}-\ket{1})\) (phase flip).

The quantities $e_1 = 1- F_1$ and $e_2 = 1 - F_2$ describe error rates for one- and two qubit
gates, and relate to the average gate fidelities $F_{1/2}$, which measures gate quality (an
exact definition follows later). Similarly, $TG_1$ and $TG_2$ denote average gate times of
one and two qubit gates, whereas $n$ specifies the number of available qubits, and $C$ is
coupling density (\ie, the average fraction of degree of connections between qubits; 100\% for
a fully meshed graph that represents physical all-to-all connectivity).

The systems characterised in Table~\ref{tab:hwdata}, albeit they only represent 
a fraction of the current variation in implementation technologies,%
\footnote{It would have been desirable to include additional vendors and approaches in our 
simulations. Yet  at the time of writing, public availability of the corresponding low-level data
is scarce, and many vendors are reluctant to publish specific values. While we rely on
vendor-reported error rates for the available data, it needs to be kept in mind that
details of how these numbers were obtained are not always clearly specified. Since (commercial) vendors might be interested in a favourable 
representation of their products, any simulations based on these numbers should be used as 
indicators, not as absolute and scientifically verified performance measures.}
exhibit widely varying characteristics that are not straightforward to translate into established
quality, performance, or scalability indicators, as they are typically considered in software 
engineering. Therefore, empirical characterisation and a generic understanding of 
the impact of imperfections on software qualities, as we address it in this paper, seems
indispensable.

\begin{table*}
    \centering
    \begin{tabular}{lllllllllll}
    \toprule
    Implementation Technology & Vendor & System & $T_{1}$ & $T_{2}$ & $F_1$ & $F_2$ & $TG_1$ & $TG_2$ & $n$ & $C$\\
    \midrule
    Superconducting Qubits & \ibmq & Kolkata & $109.90\mu s$ & $96.80\mu s$ & $99.968$\% & $98.909$\% & $35.56 ns$ & $415.37 ns$ & $27$ & 7.98\% \\
        Trapped Ions & \ionq & Aria & $10s-100s$ & $1s$& $99.95$\% & $99.6$\% & $135\mu s$ & $600\mu s$ & $21$ & 100\% \\
    Superconducting Qubits & \rigetti & Aspen M3 & $24.98\mu s$ & $28.04\mu s$ & $99.614$\% & $90.588$\% & $40 ns$ & $240 ns$ & 80 & 3.35\% \\
    \bottomrule
    \end{tabular}
    \caption{Low level hardware metrics for three commercially available QC platforms (see the text for details).}\label{tab:hwdata}
\end{table*}

We obtained the low level metrics for \ibmq system from so called \emph{FakeBackends} which are embedded in Qiskit and are based on
snapshots of their systems. For the special case of the $Z$ and $Rz$ gate, we set the respective gate time and error in our noise model to zero, since these gates are implemented virtually by all represented vendors. For circuit depth estimation, these gates are not considered either.
For \ionq the low level metrics were taken from \cite{IonqHwData}. Since the data for $T_1$ coherence has a wide range we decided on a value of $50s$ in our simulations for \ionq. Vendor \rigetti provides error data obtained via randomised benchmarking as an online 
resource~\cite{RigettiHwData}.
\ionq and \rigetti, to the best of our knowledge, do not state exactly whether their $Rz$ gates are considered in calculation of average error rates
for one qubit gates. Thus, the average error rates might be slight overestimations when compared to \ibmq. This is negligible since the two qubit gates introduce errors with higher impact.

\begin{table}
    \newcommand{\yes}{\ding{51}}
    \newcommand{\no}{\ding{55}}
    \centering
    \begin{tabular}{lllllll}
\toprule
1-Qubit Gates & X & \(\sqrt{X}\) & \(R_{x}\) &\(R_{z}\) & GPi1 & GPi2\\ 
\midrule 
\ibmq Kolkata    & \yes   & \yes   & \no  & \yes & \no  & \no  \\ 
\ionq Aria       & \no    & \no    & \no  & \yes & \yes & \yes \\ 
\rigetti AspenM3 & (\yes) & (\yes) & (\yes) & \yes & \no  & \no  \\  
\midrule
2-Qubit Gates   & \Cgate{X} & \Cgate{Z} & \Cgate{p} & XY & MS \\ 
\midrule
\ibmq Kolkata    & \yes & \no  & \no  & \no  & \no \\ 
\ionq Aria       & \no  & \no  & \no  & \no  & (\yes)\\ 
\rigetti AspenM3 & \no  & \yes & \yes & \yes & \no\\
\bottomrule
    \end{tabular}
    \caption{Native gate sets for the investigated hardware architectures. Check
    marks (\yes) denote supported; crosses (\no) denote unsupported gates.
    The parenthesised gates emphasise differences in the matrix formulation to more commonly used definitions. While we cannot 
    discuss peculiarities of gates specific to different architectures, the reproduction package supports comparative experiments based on all gate sets.
    }\label{tab:gatesets}
\end{table}

\autoref{tab:hwdata} shows error rates for quantum gates on different contemporary 
hardware approaches. All architectures are affected by
noise, which limits the achievable depth of quantum circuits, and thus the
computational power. Yet, there is no common noise pattern across systems, which
makes most statements about performance and behaviour of quantum algorithms 
impossible without taking very specific hardware details into consideration, in
stark contrast to classical machines.
As mentioned in \autoref{sec:Introduction}, for quantum error correction to work the requirement for thousands of qubits \cite{Fowler_2012} arises which will most likely not be possible on a larger scale in the near future.

\subsubsection{Connectivity}
Logical quantum algorithms usually make arbitrary (pairs of) qubits interact when
multi-qubit operations are applied. Most physical implementations of quantum systems place restrictions
on the possible interactions between pairs of qubits. One important step in translating a logical into a
physical quantum circuit is to (a) map interacting logical qubits to interconnected physical qubits,
and (b) ensure that operations performed on unconnected qubits (if placement alone cannot guarantee
this) are enabled by \enquote{moving} qubits into proximity prior to gate execution. While qubits
cannot (for most implementation technologies) be moved physically, applying
\swpgates allows us to change to logical state between two connected qubits. This allows, at the
expense of increasing circuit depths, to bring two unconnected qubits into connected positions, and then
apply a joint gate operation. For architectures that do also not natively
support logical swap gates, a replacement can be provided by
three \Cgate{X} gates, at the expense of an even larger increase in 
circuit depth.

\begin{figure}[htb]
    \centering
    \begin{subfigure}{.4\columnwidth}
        \centering
        \begin{tikzpicture}
[ncirc/.style={circle,fill=black}]

\newcommand*{\el}{.7cm}

\newcommand\addtonodelist[2]{\listadd#1{n#2}}
\newcommand\addtomarkednodes[1]{\addtonodelist{\markednodes}{#1}}

\forcsvlist{\addtomarkednodes}{null}

\def\angloffset{90}

\newcommand\setnodestyle[1]{\ifinlist{#1}{\markednodes}{\tikzset{#1/.style={circle,fill=lfd2}}}{\tikzset{#1/.style={circle,fill=black}}}}

\newcommand{\extendgrid}[4]{%
\setnodestyle{#2}%
\node (#2) at ($(#1) + (#3 \expandafter+\angloffset:\el)$) [#2] {}; \draw (#1) -- (#2);}

\node (n0) at (0,0) [ncirc] {};
\extendgrid{n0}{n1}{-45};
\extendgrid{n1}{n2}{-90};
\extendgrid{n2}{n3}{-135};
\extendgrid{n3}{n4}{-180};
\extendgrid{n4}{n5}{-225};
\extendgrid{n5}{n6}{-270};
\extendgrid{n6}{n7}{-315};
\draw (n7) -- (n0);
\extendgrid{n7}{n28}{90};
\extendgrid{n0}{n19}{90};
\extendgrid{n19}{n18}{45};
\extendgrid{n18}{n9}{90};
\extendgrid{n9}{n8}{135};
\extendgrid{n8}{n39}{180};
\extendgrid{n39}{n38}{225};
\extendgrid{n38}{n29}{270};
\draw (n29) -- (n28);
\draw (n28) -- (n19);
\extendgrid{n18}{n53}{0};
\extendgrid{n9}{n54}{0};
\draw (n54) -- (n53);
\extendgrid{n53}{n52}{315};
\extendgrid{n52}{n51}{0};
\extendgrid{n51}{n50}{45};
\extendgrid{n50}{n49}{90};
\extendgrid{n49}{n48}{135};
\extendgrid{n48}{n55}{180};
\draw (n55) -- (n54);
\extendgrid{n52}{n17}{270};
\extendgrid{n51}{n10}{270};
\draw (n17) -- (n10);
\extendgrid{n17}{n16}{225};
\draw (n1) -- (n16);
\extendgrid{n16}{n15}{270};
\draw (n2) -- (n15);
\extendgrid{n15}{n14}{315};
\extendgrid{n14}{n13}{0};
\extendgrid{n13}{n12}{45};
\extendgrid{n12}{n11}{90};
\draw (n11) -- (n10);
\extendgrid{n12}{n25}{0};
\extendgrid{n11}{n26}{0};
\draw (n26) -- (n25);
\extendgrid{n25}{n24}{315};
\extendgrid{n24}{n23}{0};
\extendgrid{n23}{n22}{45};
\extendgrid{n22}{n21}{90};
\extendgrid{n21}{n20}{135};
\extendgrid{n20}{n27}{180};
\draw (n27) -- (n26);
\extendgrid{n49}{n62}{0};
\extendgrid{n50}{n61}{0};
\draw (n62) -- (n61);
\extendgrid{n61}{n60}{315};
\draw (n60) -- (n27);
\extendgrid{n60}{n59}{0};
\draw (n59) -- (n20);
\extendgrid{n59}{n58}{45};
\extendgrid{n58}{n57}{90};
\extendgrid{n57}{n56}{135};
\extendgrid{n56}{n63}{180};
\draw (n63) -- (n62);
\extendgrid{n21}{n36}{0};
\extendgrid{n22}{n35}{0};
\draw (n36) -- (n35);
\extendgrid{n35}{n34}{315};
\extendgrid{n34}{n33}{0};
\extendgrid{n33}{n32}{45};
\extendgrid{n32}{n31}{90};
\extendgrid{n31}{n30}{135};
\extendgrid{n30}{n37}{180};
\draw (n36) -- (n37);
\extendgrid{n37}{n68}{90};
\extendgrid{n30}{n67}{90};
\draw (n68) -- (n67);
\extendgrid{n67}{n66}{45};
\extendgrid{n66}{n65}{90};
\extendgrid{n65}{n64}{135};
\extendgrid{n64}{n71}{180};
\extendgrid{n71}{n70}{225};
\draw (n70) -- (n57);
\extendgrid{n70}{n69}{270};
\draw (n69) -- (n58);
\draw (n69) -- (n68);
\extendgrid{n31}{n46}{0};
\extendgrid{n32}{n45}{0};
\draw (n45) -- (n46);
\extendgrid{n45}{n44}{315};
\extendgrid{n44}{n43}{0};
\extendgrid{n43}{n42}{45};
\extendgrid{n42}{n41}{90};
\extendgrid{n41}{n40}{135};
\extendgrid{n40}{n47}{180};
\draw (n47) -- (n46);
\extendgrid{n47}{n76}{90};
\extendgrid{n40}{n75}{90};
\draw (n76) -- (n75);
\extendgrid{n75}{n74}{45};
\extendgrid{n74}{n73}{90};
\extendgrid{n73}{n72}{135};
\extendgrid{n72}{n79}{180};
\extendgrid{n79}{n78}{225};
\draw (n78) -- (n65);
\extendgrid{n78}{n77}{270};
\draw (n77) -- (n66);
\draw (n77) -- (n76);

\end{tikzpicture}
        \caption{\rigetti Aspen M3}
        \label{fig:aspenm3_topo}
    \end{subfigure}
    \begin{subfigure}{.55\columnwidth}
        \begin{subfigure}{\textwidth}
        \centering
            \begin{tikzpicture}
[ncirc/.style={circle,fill=black}]

\newcommand*{\el}{.7cm}

\newcommand\addtonodelist[2]{\listadd#1{n#2}}
\newcommand\addtomarkednodes[1]{\addtonodelist{\markednodes}{#1}}

\forcsvlist{\addtomarkednodes}{null}

\def\angloffset{30}

\newcommand\setnodestyle[1]{\ifinlist{#1}{\markednodes}{\tikzset{#1/.style={circle,fill=lfd2}}}{\tikzset{#1/.style={circle,fill=black}}}}

\newcommand{\extendgrid}[4]{%
\setnodestyle{#2}%
\node (#2) at ($(#1) + (#3 \expandafter+\angloffset:\el)$) [#2] {}; \draw (#1) -- (#2);}

\node (n0) at (0,0) [ncirc] {};

\extendgrid{n0}{n1}{300};
\extendgrid{n1}{n2}{240};
\extendgrid{n2}{n3}{240};
\extendgrid{n3}{n5}{300};
\extendgrid{n5}{n8}{300};
\extendgrid{n8}{n9}{240};
\extendgrid{n8}{n11}{0};
\extendgrid{n11}{n14}{0};
\extendgrid{n14}{n13}{60};
\extendgrid{n13}{n12}{60};
\extendgrid{n12}{n10}{120};
\extendgrid{n10}{n7}{120};
\extendgrid{n7}{n6}{60};
\extendgrid{n7}{n4}{180};
\draw (n4) -- (n1);
\extendgrid{n14}{n16}{300};
\extendgrid{n16}{n19}{300};
\extendgrid{n19}{n20}{240};
\extendgrid{n19}{n22}{0};
\extendgrid{n22}{n25}{0};
\extendgrid{n25}{n26}{300};
\extendgrid{n25}{n24}{60};
\extendgrid{n24}{n23}{60};
\extendgrid{n23}{n21}{120};
\extendgrid{n21}{n18}{120};
\extendgrid{n18}{n17}{60};
\extendgrid{n18}{n15}{180};
\draw (n15) -- (n12);

\end{tikzpicture}
            \caption{\ibmq Kolkata}
            \label{fig:kalkata_topo}
        \end{subfigure}
        \begin{subfigure}{\textwidth}
            \centering
            \tikzset{ncirc/.style={circle,fill=#1}}
\begin{tikzpicture}[ncircc/.style={circle,fill=black}]
\pgfmathsetmacro{\numnodes}{21};
\foreach \i in {1,2,...,\numnodes}{
    \tikzmath{
        \angl =  \i * (360 / \numnodes);
    };
    \node (n\i) at (\angl:4.5cm) [ncirc=black] {};
    \foreach \j in {1,...,\i}{
        \pgfmathsetmacro{\col}{ifthenelse(\i!=\numnodes,"gray","black")};
        \pgfmathsetmacro{\linesize}{ifthenelse(\i!=\numnodes,".3pt","2pt")};
        \draw [\col, line width=\linesize] (n\i) -- (n\j);
    }
}
\end{tikzpicture}
            \caption{\ionq Aria}
            \label{fig:ionq_topo}
        \end{subfigure}    
    \end{subfigure}
    
    \caption{Hardware connectivity graphs from various commercial vendors considered in the numerical 
    simulations. The implementations of vendors \rigetti and \ibmq both have a grid like structure,
    and differ slightly in their connectivity structure. The \ionq architecture (and other trapped
    ion systems) features full connectivity between qubits.}
    \label{fig:toplogies}
\end{figure}
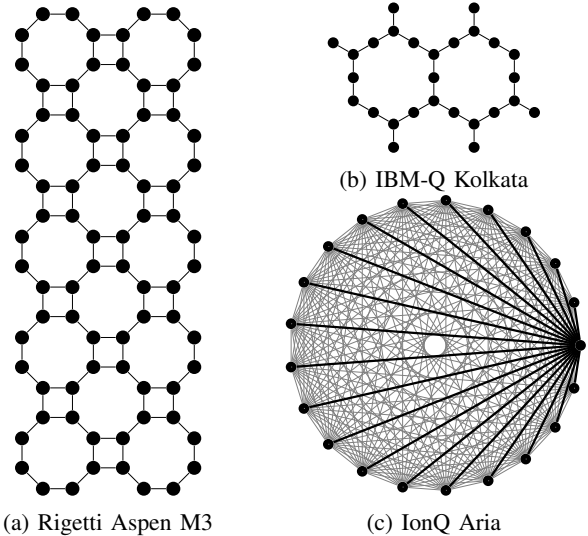

\autoref{fig:toplogies} compares topologies for some of the major available quantum
architectures. We illustrate their influence on
algorithms in Section~\ref{sec:connectivity_gates}.

\subsubsection{Gate Sets}
The set of elementary quantum gates varies considerably with implementation technology.
Universal quantum computation can be achieved with many different choices. While the
theoretical capabilities of each set are identical, the practical behaviour of gates
under the influence of noise may vary distinctly. Executing identical algorithms
on different hardware does therefore not only influence computation times (as is familiar
from classical computing), but is also affected by different influence of noise.
\autoref{tab:gatesets} illustrates elementary gate sets for the subject architectures.

\begin{figure*}
    \centering
        \begin{tikzpicture}
\yquantdefinegate{Rzpi}{
    qubit a;
    
}

\tikzset{
    rotgate/.style={fill=lfd2,draw=none}
}

\begin{yquant}[operators/every h/.append style={fill=lfd4,draw=none},
operators/every x/.append style={fill=lfd3,draw=none},
every control line/.append style={draw=lfd3, thick},
every positive control/.append style={fill=lfd3},
operators/every zz/.append style={fill=lfd3}]
qubit q[3];

h q[0]; h q[1]; h q[2];
x q[1]; x q[2];
[name=zz1] zz (q[0], q[1], q[2]);
x q[1]; x q[2];
align q; h q[0]; h q[1]; h q[2];
x q[0]; x q[1]; x q[2];
zz (q[0], q[1], q[2]);
x q[0]; x q[1]; x q[2];
h q[0]; h q[1]; h q[2];
x q[1]; x q[2];
\end{yquant}
\node (zz1bot) at ($(zz1-0) + (0,-3.2em)$){};
\node [draw, dashed, fit=(zz1-0) (zz1bot)] {};
\begin{scope}[shift={(7,0)}]
\begin{yquant*}[operators/every x/.append style={fill=lfd3,draw=none},
every control line/.append style={draw=lfd3, thick},
every positive control/.append style={fill=lfd3}]
qubit {} q[3];

[name=left]
[rotgate]box {$R_{z}^{\frac{\pi}{4}}$} q[1]; [rotgate]box {$R_{z}^{\frac{\pi}{4}}$} q[2];
x q[2] | q[1];
[rotgate]box {$R_{z}^{-\frac{\pi}{4}}$} q[2];
x q[2] | q[1];
[rotgate]box {$R_{z}^{\frac{\pi}{4}}$} q[2]; x q[0] | q[1];
[name=top]
[rotgate]box {$R_{z}^{-\frac{\pi}{4}}$} q[0]; 
x q[0] | q[1];
x q[1] | q[0];
x q[0] | q[1];
x q[2] | q[1];
[rotgate]box {$R_{z}^{\frac{\pi}{4}}$} q[2]; 
x q[2] | q[1];
[rotgate]box {$R_{z}^{-\frac{\pi}{4}}$} q[2]; x q[1] | q[0];
align q; [rotgate]box {$R_{z}^{\frac{\pi}{4}}$} q[1]; 
x q[2] | q[1];
[rotgate]box {$R_{z}^{-\frac{\pi}{4}}$} q[2]; 
[name=rightbot]
x q[2] | q[1];
\end{yquant*}
\end{scope}
\node (zztransp) [draw, dashed, inner sep=.7em, fit=(left-0) (top-0) (rightbot-0)] {};
\draw [-Stealth] (zz1-0) ++(0,.5em) -- ($(zz1-0) + (0,2.5em)$) -| (zztransp);
\node at ($(zz1-0.north)+(0,2.5em)!.45!(zztransp.north)+(0,1.5em)$) {transpile};
\end{tikzpicture}
    \caption{Effect of transpiling a gate in a logical Grover circuit (left) to IBM-Q 
    Kolkata hardware (right, dotted frame). Even a seemingly small component like a 
    (multi-) controlled Z (\(\Cgate{Z}\)) gate can introduce significant increase in circuit depth.}\label{fig:circ_vigo_3}
\end{figure*}
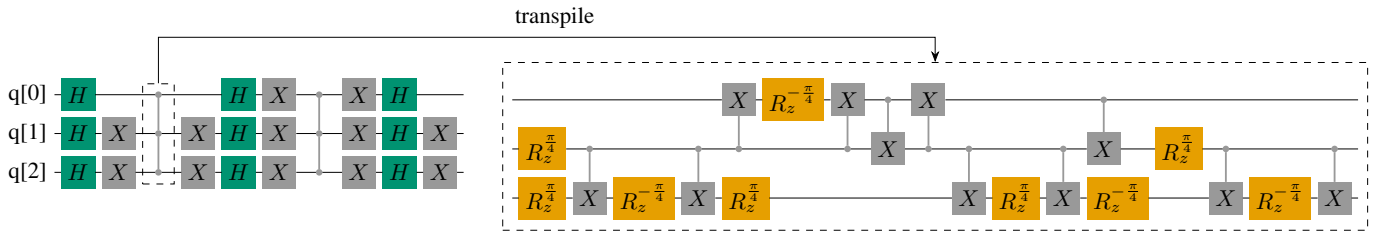

\subsection{Subject Algorithms}
We have chosen three canonical, yet substantially different algorithms to 
study the impact of noise and imperfections: Grover search, quantum Fourier
transform, and variational quantum circuits. It is possible to prove
speedups over their classical counterparts for the first two algorithms,
albeit these only materialise for perfect, error-corrected quantum systems.
Their scalability in terms of circuit depth growth with increasing input size is distinctly different.

The third class, variational quantum circuits, is speculated to exhibit speedups over classical approaches
under some credible assumptions, and are particularly well suited for
NISQ hardware and empirical experiments, as they allow for very shallow circuits. Yet,
practically relevant speedups have still failed to materialise on a wider front in current systems.

\subsubsection{Grover Search}
Grover's algorithms allows for finding specific elements in an unstructured search space.
Simply put, the algorithm iteratively repeats two sub-circuits: An oracle to mark the desired 
element in a search space, and a rotation in a two dimensional plane. For inputs of
\(n\) qubits, the required number of iterations scales with $\mathcal{O}
(\sqrt{2^n})$~\cite{nielsen_chuang_2010}, which provides a quadratic speedup compared to the
best classical search algorithms---yet, this speedup is relative to exponential growth. Since the 
algorithm matches a wide class of application problems, it can seem tantalizing to seek 
\enquote{free} quantum improvements by deploying the algorithm as drop-in search replacement in
existing scenarios. However, there are some pitfalls to consider: Grover search does usually not, 
despite common perception, query an actual \enquote{physical} database encoded in
quantum states,%
\footnote{While it would be possible \emph{in principle} to apply Grover 
search on top of quantum random access memory (QRAM), this would result in a 
quadratic speedup for a search task on an exponentially
growing search space, which is usually irrelevant industrial settings.
Other data loading alternatives exhibit similar difficulties.
Approximate encoding techniques~\cite{Nakaji:2022}, together with shallow variants
of Grover~\cite{LiuJ:2021,Brianski:2021}, or improvements in the amplitude 
amplification process~\cite{Tezuka:2022} might lead to fruition, but underline
that judging non-functional software characteristics is impossible without
accounting for technical and physical details that can be ignored in classical
approaches. Given that QRAM is invariably harder to 
manufacture than classical RAM, a scenario where the former can fully
replace the latter seems hardly credible.}
but evaluates an efficiently computable function \(f\) that acts as
predicate to identify one or more optimal elements in a search space.
This implies costs (especially in terms of circuit depth)
to \emph{implement} this target function using quantum operators, which may
be non-trivial~\cite{Qiskit-Textbook}.

Many practical applications are either interested in average case complexity, or enjoy
some structure in their search space that can be (also heuristically)
used to speed up processing, which places considerable limitations on 
practical utility. Replacing classical primitives with Grover search
is a commonly used pattern (\eg in database research~\cite{Zajac2022,Amellal2018Springer}) in efforts
to utilise quantum computing, and is sometimes backed up by evaluations on small-scale NISQ 
machine. We study the limited utility of this approach in Sec.~\ref{sec:simulations}.

By determining the probability of measuring the desired element of the search space
as output, we can associate a success probability with runs of Grover search. Our implementation is inspired by ~\cite{figgatt2017complete,Qiskit-Textbook}.

\subsubsection{Quantum Fourier Transform}
The quantum Fourier transform is as a building block for many quantum algorithms,
most notably Shor's factoring algorithm. It is a computational analogue of the (discrete)
classical Fourier transformation, albeit there are pronounced differences in 
obtaining the results, as the probability of reading a Fourier coefficient is related
to its magnitude, and applications that require access to the full transformation do
not benefit from quantum advantage. Yet, QFT requires exponentially fewer
operations than FFT. 

\subsubsection{Variational Quantum Circuits}\label{sec:var_explanation}
The algorithmic family of variational quantum algorithms comprises circuits 
that contain gates whose properties are controlled by a tunable parameter.
After feeding an input state through a circuit and measuring the output, the parameter
settings are adjusted in a training process---not unlike classical machine learning---,
and the process is repeated, until some desired target function is approximated with
sufficient quality. The approach is versatile and well suited for experimentation on
NISQ machines, particularly because of controllable circuit depth.

An example for a variational circuit (as we employ it in the below experiments)
is shown in Fig.~\ref{fig:variational_circ}. 
For data encoding, which comprises the left part of our circuit, we use a similar construction as in Ref.~\cite{CircuitLearning}.
The parameterized part of our quantum circuit is one of several building blocks for variational quantum circuits,
the authors of ~\cite{sim2019expressibility} investigate in their work. While the variational part of our circuit could be repeated several times, we stick to one single parameterized layer in our experiments.

For our simulations we train the variational circuit to mimic the behaviour of the function $f(x)=x^2$, extending an example of~\cite{CircuitLearning} for different noise levels. We will now quickly summarize the procedure:
We call the parameters  of the circuit $\theta_i \in \bm{\theta}$, all of which start at $\theta_i=0$
as initial parameter values. For simplicity we choose to measure only the qubit $\ket{q_1}$ in the $Z$ basis and
subsequently extract the expectation value of the observable $\langle M_j \rangle_{\bm{\theta}} \in [-1,1]$ for
the training input $x_j$ and the current parameters.
For sample based approaches the approximation of $\langle M_j \rangle_{\bm{\theta}}$ requires a reasonable choice of circuit estimations, that is, the sample number. Based on the quadratic loss function for iteration $k$,
\begin{equation}
L\left(x_j,\bm{\theta}^{(k)}\right) = \frac{1}{2}\left(\langle M_j \rangle_{\bm{\theta}^{(k)}} - 
f(x_j)\right)^2,
\end{equation}
we update \(\bm\theta\) in every iteration using gradient descent 
\begin{equation}
    \bm{\theta}^{(k+1)} = \bm{\theta}^{(k)} - \eta \frac{\partial L}{\partial \bm{\theta}}.
\end{equation}
We calculate the derivative of the loss function using the parameter shift rule~\cite{CircuitLearning}
\begin{equation}
\begin{split}
    &\frac{\partial L}{\partial \theta_i} =\\
    &\frac{1}{2}(\langle M_j \rangle_{\bm{\theta}^{(k)}} - f(x_j))
    \left( \langle M_j \rangle_{\bm{\theta}^{(k)}+\vec{e}_i \frac{\pi}{2}} - 
            \langle M \rangle_{\bm{\theta}^{(k)}-\vec{e}_i \frac{\pi}{2}} \right),
\end{split}
\end{equation}
where $\vec{e}_i$ is the a unit vector for component \(i\). After carrying out the partial
gradient calculation for every parameter, the accumulated gradient
$\frac{\partial L}{\partial\bm{\theta}}$ is used in the parameter update.

Our simulations fix the number of training iterations at 100, and take uniformly spaced
samples $x_j \in [-1,1]$, which are permuted with a (fixed) random seed. Results
are therefore comparable between architectures and noise variants.

\section{Modelling Noise and Imperfections}\label{sec:modelling}
In the following, we outline the theoretical concepts necessary to describe and
understand how noise (\ie, the influence of uncontrollable external factors) 
impacts quantum calculations. We aim at an exposition that is accessible to
software engineers without deeper involvement in quantum physics, yet sufficiently
accurate to paint a realistic picture that allows for drawing reliable conclusions.

\subsection{Mixed States and Density Operators}
The notion of quantum states in software engineering usually refers to \emph{pure states}
$\ket{\phi}\in \mathbbm{C}^{2^n}$, that is, states represented by a unit vector in a $2^n$
dimensional complex vector (Hilbert) space. Sometimes, however, it is not possible to obtain
full knowledge of the state of a system. Consider, for example, the case where we take an
(educated) guess whether some external influence that is not under our control (in other words,
noise) flipped qubit $x_i$ or not. In this scenario, the system is in state $\ket{x_i}$ with
probability $p$ or in state $\ket{\neg x_i}$ with probability $1-p$. Generalising this concept delivers a probability distribution  $\left(\begin{smallmatrix}\ket{\phi_1} & \ket{\phi_2} 
& \dots \\ p_1 & p_2 & \dots\end{smallmatrix}\right)$ of different possible system states.
If a system follows such a distribution it is said to be in a \emph{mixed state}.

Mixed states contain \emph{two} probabilistic aspects: (a) Stochastic outcomes
resulting from measurements have their origin in the very properties of
quantum theory. (b) Purely classical uncertainty about the state that arises from a
\emph{lack of knowledge} of external confounding factors (noise).

When influences beyond our active control modify a quantum state, we need to express the
(classical) uncertainty arising from the scenario with classical probabilities $p_i$. 
One convenient way to express mixed states is the density matrix formalism~\cite{nielsen_chuang_2010},
which describes a \enquote{collection} of quantum states \(\ket{\phi_i}\) mixed up with
classical probabilities \(p_{i}\) as
\begin{equation}
    \rho = \sum_i p_i \ket{\phi_i}\bra{\phi_i},
\end{equation}
where \(\bra{\phi}\) denotes, for finite-dimensional systems, the conjugate transpose of \(\ket{\phi}\),
turning the object \(\ket{\phi}\bra{\phi}\) into a matrix. Application of a unitary operator $U$,
which represents a computational step, to a mixed state $\rho$ is described by
$\rho\rightarrow U \rho U^\dagger$. In the following, transformations of the density matrix
will be used to describe evolution of quantum states suspect to noise.

\subsection{Dynamics of Noisy Quantum Programs}
\label{sec:dynamicsNoisyQuantum}
The evolution of a closed quantum system is described by unitary operations. A noisy system on 
the contrary is open to an external environment (source of noise). The trick to model open noisy 
systems is to include the environment, such that one ends 
up with a bigger but closed quantum system. Let $\rho$ be the state of an open quantum system of 
interest, which we will call the principal system. Now, we combine $\rho$ with the state of its 
environment $\rho_{\text{env}}$. The new system $\rho \otimes \rho_{\text{env}}$ is closed and 
its evolution $U (\rho \otimes \rho_{\text{env}}) U^\dagger$ can be described by a unitary 
operator $U$. Tracing out the environment reveals the evolution of $\rho$ under noise: 
$\mathcal{E}(\rho) = \tr_{\text{env}}(U (\rho \otimes \rho_{\text{env}}) U^\dagger)$. The partial 
trace is a tool in the density matrix formalism to discard certain parts of a quantum mechanical 
systems, for more information we refer to \cite{nielsen_chuang_2010}. Note that the \emph{quantum 
operator} $\mathcal{E}(\rho)$ is not necessarily unitary anymore. Let $\mathcal{B}_e = 
\{\ket{e_k}\}_k$ be a basis of the environment. Now, if the environment is measured in 
$\mathcal{B}_e$ after the time evolution, then the outcome determines the state of the principal 
system. We end up with a random distribution of states for the principal system depending on the 
measurement. The effect the environment had on $\rho$ when the outcome $k$ occurred can be 
described by an operator $E_k$~\cite{nielsen_chuang_2010,Kraus:1983}, leading to a mixed state description
\begin{equation}
    \rho \mapsto \sum_i E_i \rho E_i^{\dagger}, ~ \sum_i E_i^\dagger E_i = 1
    \label{eq:KrausFormulation}
\end{equation}

\subsection{Fidelity}
In this work we mainly focus on the effect of noise when performing computations, that is gate 
errors. Consider a quantum operation $\mathcal{E}(\rho)$ describing the noise impact on 
$\rho$. Under the influence of noise, a pure state $\rho = \ketbra{\psi}$ evolves to $\mathcal{E(\rho)} = 
\sum_i p_i \ketbra{\psi_i}$. We can now calculate $\ev{\mathcal{E}(\rho)}{\psi} = \sum_i p_i \braket{\psi}{\psi_i} 
\braket{\psi_i}{\psi} = \sum_i p_i \abs{\braket{\psi}{\psi_i}}^2$, which measures the overlap between $\ket{\psi}$ 
and $\mathcal{E}(\rho)$. This can be seen as a measure of how much information is preserved under 
noise. In a perfect noiseless environment $\mathcal{E}(\rho) = \ketbra{\psi}{\psi} = \rho$, 
preserving all the information, respectively $\ev{\mathcal{E}(\rho)}{\psi} = \abs{\braket{\psi}
{\psi}}^2 = 1$. On the other hand, the more $\mathcal{E}(\rho)$ turns $\ket{\psi}$ in the 
direction of an orthogonal state $\ket{\psi^\bot}$ the more information gets lost and 
$\ev{\mathcal{E}(\rho)}{\psi}$ approaches $0$, as \(\mathcal{E}(\rho)\) goes to 
\(\ketbra{\psi^\bot}\). 

The measure $F=\ev{\mathcal{E}(\rho)}{\psi}$ is commonly denoted as the
\emph{fidelity};\footnote{Different definitions of fidelity are given in the literature;
as the general characteristics are identical, we only consider one variant in this paper..} 
it is useful to judge how much the result of a noisy quantum computer deviates form the result of a perfect machine.

The deviation of imperfect quantum gates (or other components) from a perfect 
implementation is not directly captured by fidelity: Gates do not operate
on a fixed input state. Their degree of deviation from a perfect gate is
state-dependent, so determining gate fidelity for a single state is insufficient
to characterise quality. Instead, the \emph{average 
fidelity} describes the mean over individual gate fidelities for all quantum
states.\footnote{How to compute the average of a desired quantity over all possible quantum states
of a system is an interesting problem on its own that we cannot
discuss in detail; see, for instance, Ref.~\cite{simpleNielsen,gilchrist2005Distance} for more information.}
The values shown in~\autoref{tab:hwdata} represent
average fidelities.

\subsection{Noise Models}
Having introduced the general modelling principles for noisy quantum systems, we can now
describe specific types of noise that represent typical physical imperfections.

\subsubsection{Bit Flips}
A probabilistic qubit flip~\cite{nielsen_chuang_2010} is given by 
\begin{equation}
    \rho \mapsto  (1-p) \mathbbm{1} \rho\mathbbm{1}^{\dagger} + p X \rho X^\dagger.
    \label{eq:bitFlipChannel}
\end{equation}
The operation applies the Pauli $X$ gate (bit flip) to the one qubit system $\rho$ with 
probability $p$, and else leaves the state as is. Note how the above equation is one way to choose
the operators $E_k$ in \eqref{eq:KrausFormulation}. Similarly, the phase flip error (Pauli $Z$ gate)
and the bit-phase flip error (Pauli $Y$ gate) can be constructed by replacing  $X$ by $Z$ or $Y$ in 
Eq.~(\ref{eq:bitFlipChannel}).

\subsubsection{Depolarisation}
One commonly used error in simulation is the completely depolarising operator on one
qubit which randomly applies the Pauli operators $X,Y,Z$ \cite{ekert,nielsen_chuang_2010} 

\begin{equation}
\begin{aligned}
\label{eq:depolChannelKraus}
\rho \:\: \mapsto &\:\:(1-p) \mathbbm{1} \rho \mathbbm{1}^\dagger + \\
             &\:\:p\frac{1}{4}\left(\mathbbm{1} \rho \mathbbm{1}^\dagger +  X\rho X^\dagger+Y\rho Y^\dagger+Z\rho Z^\dagger\right), 
\end{aligned}
\end{equation}
with a certain probability, and else leaves the qubit as is. A quick calculation reveals 
that \eqref{eq:depolChannelKraus} equals $\rho \mapsto (1 - p) \rho + p \frac{1}{2} 
\mathbbm{1}$, where $\frac{1}{2}\mathbbm{1}$ is the 
density representing the state of a system being in every basis state with equal 
probability. Hence, the system either stays intact or all information gets destroyed with 
probability $p$. For a $n$ qubit system we get\cite{nielsen_chuang_2010}:
\begin{equation}
    \rho \mapsto(1-p)\rho + p \frac{1}{2^n}\mathbbm{1} 
    \label{eq:depolSimpleForm}
\end{equation}

\subsubsection{Thermal Relaxation}
The thermal-relaxation error models the decoherence of quantum system over time. The derivation
of the associated quantum operator is significantly less straight forward, so we refer to 
Refs.~\cite{Georgopoulos2021ModelingAS,blank2020quantum} for a derivation.

\subsubsection{Hardware-Matched Composite Noise}
We close by incorporating hardware metrics into a noise model 
(limited to stochastic gate noise, and ignoring measurements, state preparation, idle noise and coherent error models).
We construct three noise models from the data in \autoref{tab:hwdata}.\footnote{Our approach is a simplified version of the integrated error models for \ibmq \enquote{FakeBackends} provided by Qiskit~\cite{Qiskit}.}

The idea is to use a composite error consisting of thermal relaxation depolarisation
for every gate, such that average gate fidelities of the real hardware match the model \cite{qiskitdeverror}.
For our simplified version we only distinguish between one and two qubit gates, but do not
introduce per-gate errors, or errors depending on individual qubit quality. While this
renders the model less accurate than, for instance, \ibmq \enquote{FakeBackends},
it allows us to apply it to hardware for which detailed quality data are not publicly available.
The construction goes as follows: We define an error operator $\mathcal{E} = \mathcal{E}_D \circ 
\mathcal{E}_R$, combining both the depolarising error $\mathcal{E}_D$ \eqref{eq:depolChannelKraus}
and the thermal relaxation error $\mathcal{E}_R$. The fidelity of $\mathcal{E}_D$ is given 
by~\cite{MagesanDepolarizingFidelity2011}
\begin{equation}
    F_{D} = 1 - p(1-2^{-n}),
    \label{eq:fidDepolValue}
\end{equation}
and the fidelity for thermal relaxation channel $\mathcal{E}_R$ can be calculated using its parameters, namely $T_1$, $T_2$ and the gate time $T_G$. 
We tune our model to match target fidelities $F_{\text{targ}}$ found in the literature. Using \eqref{eq:depolSimpleForm}, the composition of depolarizing and thermal relaxation channel is $\mathcal{E}_{D} \circ \mathcal{E}_{R} = (1-p) \mathcal{E}_R(\rho) + p \frac{1}{2^n}\mathbbm{1}$. The fidelity denotes the overlap with a pure reference state $\ket{\psi}$,
\begin{equation}
\begin{split}
    F(\mathcal{E}_{D} \circ \mathcal{E}_{R})&=\expval{(1-p)\mathcal{E}_R(\rho)+p\frac{\mathbbm{1}}{2^n}}{\psi}\\
    &=(1-p) \expval{\mathcal{E}_R(\rho)}{\psi} + p \expval{\frac{\mathbbm{1}}{2^n}}{\psi},
\end{split}\label{eq:braketFidelity}
\end{equation}
where we use the linearity of the inner product in the second argument. The left hand side of
Eq.~(\ref{eq:braketFidelity}) is $(1-p)F(\mathcal{E}_R)$, while the right hand side is $p F(\mathcal{E}_D|_{p=1}) =p 2^{-n}$, which corresponds to
the fidelity of the depolarising channel evaluated at $p=1$.

We can express the target fidelity dependent on $p$, and solve
\begin{equation}
        \begin{split}
        F_{\text{arg}} &= F(\mathcal{E}_D \circ \mathcal{E}_R)
        =(1-p)F_R + p 2^{-n}\\
        \Leftrightarrow p &= \frac{F_{R}-F_{targ}}{F_{R}- 2^{-n}}.
        \end{split}
\end{equation}
This ensures the composition matches the vendor target fidelity $F_{\text{targ}}$
given in \autoref{tab:hwdata}.

\subsection{Measurement and Sampling}
It is textbook knowledge that measuring quantum states results in probabilistic
outcomes; a natural question that needs to be addressed to characterise algorithmic
performance is how many samples are required to achieve acceptable trust in outcomes.
Given an error margin \(\epsilon\) that we are willing to accept, and a desired
confidence $\delta$ for the sampled probabilities to fall within this margin of error,
a lower bound on the required number of samples \(s\) can be determined by invoking the
variant \(s \geq \frac{1}{2\epsilon^{2}}\operatorname{log}(2/\delta)\) of the Höffding inequality (see, \eg,~\cite{PashayanQuasiprobs}),
to arrive at meaningful statements.
Note that the numerical experiments considered in Sec.~\ref{sec:simulations} simulate
the complete density matrix, from which \emph{exact} measurement statistics can be
deterministically extracted. Therefore, no sampling noise as it would arise for
real hardware is contained in the plots.
Accessing the density matrix on real hardware is possible, but requires an
(experimentally involved) quantum state tomography that measures a complete set 
of observables whose expectation values determine the density operator (a number of 
measurements exponential in the system size is required, albeit less costly approximations
are possible~\cite{Cramer:2010}). Consequently, sampling is unavoidable to characterise
quantum algorithms on real systems, and the above considerations guide software engineers
on what temporal overheads to expect.

\section{Numerical Simulations}\label{sec:simulations}
We now commence with illustrating the concrete effects of the 
various modes of imperfection on the subject algorithms, and show how they 
crucially affect many algorithmic properties that are directly relevant
for software engineering. We deliberately base our considerations
on two seminal, very well understood algorithms despite their known
non-usefulness on NISQ hardware, as it allows us to focus on the influence
of noise instead of having to consider peculiarities of the subject algorithms.
This is important to approach the topic from a tangible, concrete software
engineering perspective that links the influence of noise with what computer
scientists are well acquainted with: The performance analysis of algorithms.

\begin{figure}[htbp]
\centering\begin{tikzpicture}[every node/.style={semithick,align=center,
                    text centered,text width=3.5cm,minimum height={0.75cm},font=\small},
                    node distance=5mm and 1cm]

    \node[draw,dashed] (config) {Simulation Parameters};
    \node[draw,below right = of config] (base) {Circuit Generation};
    \node[draw,below = of config] (noise) {Noise Generation};
    \node[draw,below = of noise] (density) {Density Matrix Simulator};
    \node[draw,right = of density] (map) {Translation \& Mapping};
    \node[draw,dashed,below = of density] (measure) {Data Extraction};

    \node[draw=gray,dotted,fit=(map) (density) (measure)] (var){};
    \node[draw=gray,color=gray,below = of var] (classic) {Classical Optimiser};
    
    \draw[-Stealth] (config) -| (base);f
    \draw[-Stealth] (config) -- (noise);
    \draw[-Stealth] (noise) -- (density);
    \draw[-Stealth] (base) -- (map);
    \draw[-Stealth] (map) -- (density);
    \draw[-Stealth] (density) -- (measure);
    \draw[Stealth-Stealth,draw=gray,color=gray] (var) -- node [text width=4cm,anchor=west] {$\theta^{(k+1)} = \theta^{(k)} - \eta \frac{\partial L}{\partial \theta}$} node [text width = 1cm, anchor=east,color=gray] {$\circlearrowleft$} (classic) ;

    \node[below = of map,color = gray] (des) {Repeat for Variational Circuits};
\end{tikzpicture}
\caption{Simulation procedure. Gray parts only apply to variational
    circuits with iterative parameter
    optimisation.}\label{fig:overview_fig}
\end{figure}
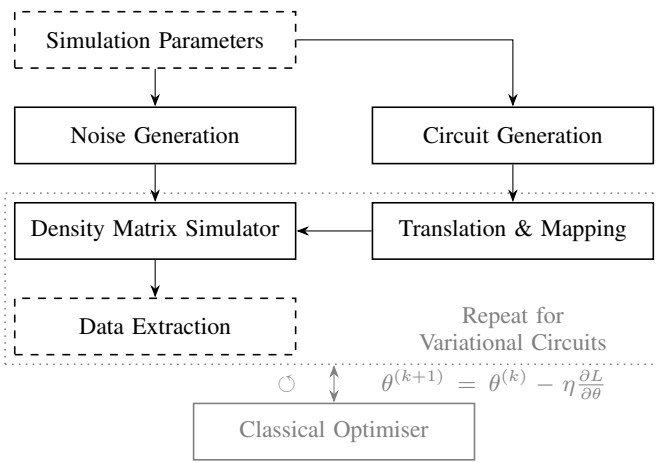

\autoref{fig:overview_fig} summarises the simulation procedure as 
implemented in the reproduction package (available on the \suppweb). 
First, a quantum circuit is generated, together with a noise model characterised by type of
noise and strength, as specified by the input parameters. Then, the logical circuit is translated
into a physical representation for one of the supported hardware platforms, which comprises a
user-customisable topology and gate set, and does not necessarily correspond to a real physical
platform, albeit we focus on the platforms characterised in~\autoref{tab:hwdata}. This allows users to consider tailored designs that match the requirements of problems of interest, and can guide
(co-)development of quantum hardware. The reproduction package extends
standard means provided by Qiskit with gate sets for \ionq and \rigetti, as well
as methods for mapping base circuits onto these.\footnote{Our results are obtained from the Qiskit density matrix backend, which provides
complete and accurate results, and is not restricted with respect to simulating noise. We use
a standard gradient descent optimiser to iteratively improve parameters for variational algorithms.}

\subsection{Noise and Scalability}
\begin{figure}[htbp]
    \includegraphics{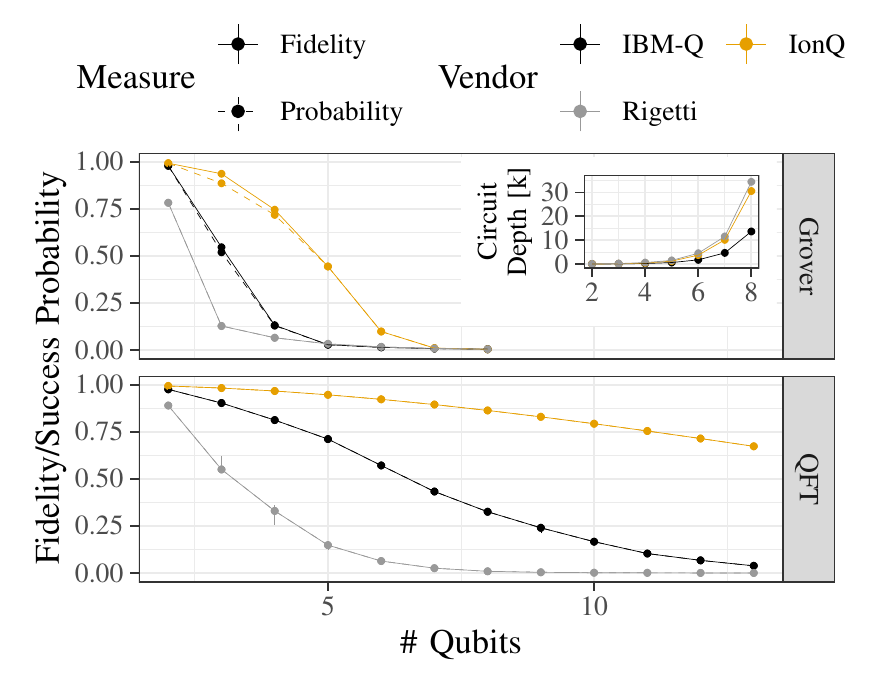}\vspace*{-1em}
    \caption{Success probability and fidelity (relative to a noise-free 
    perfect execution) for Grover Search and QFT. 
    }\label{fig:grover+qft}
\end{figure}

We start by considering scalability with respect to input size and 
noise strength. While it is well known that Grover Search and QFT can not be realistically 
deployed on NISQ machines, it is instructive to 
see how detrimental effects of noise on their performance are. \autoref{fig:grover+qft} simulates
success probabilities using the inherent native machine noise for both algorithms.

For all hardware architectures, the success probability quickly 
drops; Grover search does not produce valid solutions for more than six qubits on
any architecture. This clearly indicates that experimental evaluations that are intended 
to show the practical functionality Grover-based approaches carry little merit; the 
principal functioning of Grover's algorithm is very well understood, but it is hardly
possible to extrapolate any meaningful statements from such measurements. 
It is also instructive to consider how the slight differences in fidelity (recall~\autoref{tab:gatesets}) lead to comparatively large differences in success probabilities
for the different vendor platforms.

As~\autoref{fig:grover+qft} shows, success probability and state fidelity
are essentially identical. While it is possible to define a \enquote{successful} 
target state in the computational basis for Grover and QFT, this does not necessarily
hold for other approaches, and does also not extend to approximation cases where closeness
to an ideal state is sought. Fidelity (as a continuous measure of closeness) is a 
useful replacement for success probability.

The inset in the top part of \autoref{fig:grover+qft} illustrates the increase 
in circuit depth with increasing input size; \autoref{fig:gatesets} further 
below provides the same information for QFT (for now, only consider the 
elements for native connectivity). Circuit depths
for Grover increase exponentially, and exceed values of 1000 for more than
6 qubits, resulting in zero success probability. The increase for QFT is more
relaxed, with depths of around 150 for all architectures at 11 qubits of input. 
The differences in success probability highlight circuit depth as key performance
(and feasibility) indicator, which should therefore be at the core interest
of software engineers.

\subsection{Noise Variants}
\begin{figure}[htbp]
    \newcommand{\R}[3]{R_{#1}^{#3}}
%
\yquantset{register/default name=$\ket{\reg_{\idx}}$}
\begin{small}\begin{tikzpicture}
\yquantdefinegate{Rzpi}{
    qubit a;
    
}

\tikzset{
    rotgatez/.style={fill=lfd2,draw=none},
    rotgatey/.style={fill=lfd4,draw=none}
}

\begin{yquant}[operators/every h/.append style={fill=lfd4,draw=none},
operators/every x/.append style={fill=lfd3,draw=none},
every control line/.append style={draw=lfd3, thick},
every positive control/.append style={fill=lfd3},
operators/every zz/.append style={fill=lfd3}]

\begin{scope}[shift={(-0.01,0)}]
qubit q[4];
[name=ol]
[rotgatey]box {$R^\xi_{y}$} q[0];
[rotgatey]box {$R^\xi_{y}$} q[1];
[rotgatey]box {$R^\xi_{y}$} q[2];
[name=ur]
[rotgatey]box {$R^\xi_{y}$} q[3];
\end{scope}
\begin{scope}[shift={(0.5,0)}]
[name=ol2]
[rotgatey]box {$\R{y}{\theta}{0}$} q[0];
[rotgatey]box {$\R{y}{\theta}{1}$} q[1];
[rotgatey]box {$\R{y}{\theta}{2}$} q[2];
[rotgatey]box {$\R{y}{\theta}{3}$} q[3];

[rotgatez]box {$\R{z}{\theta}{4}$} q[0];
[rotgatez]box {$\R{z}{\theta}{5}$} q[1];
[rotgatez]box {$\R{z}{\theta}{6}$} q[2];
[name=theta7]
[rotgatez]box {$\R{z}{\theta}{7}$} q[3];

x q[0] | q[1];
x q[2] | q[3];
[rotgatey]box {$\R{y}{\theta}{8}$} q[1];
[rotgatey]box {$\R{y}{\theta}{9}$} q[2];
[rotgatez]box {$\R{z}{\theta}{10}$} q[1];
[rotgatez]box {$\R{z}{\theta}{11}$} q[2];
[name=r2]
x q[1] | q[2];

\node (enc) [draw, dashed, inner sep=.7em, fit=(ol) (ur) ] {};
\node (var) [draw, dashed, inner sep=.7em, fit=(ol2) (r2) (theta7) ] {};
\node (xj) at ($(enc.north)+(0,1.5em)$) {$x_j$};


\node at ($(var.north)+(0,1.5em)$) {$\bm{\theta}^{(k+1)} = \bm{\theta}^{(k)} - \eta \frac{\partial L}{\partial \bm{\theta}}$};

\end{scope}
\begin{scope}[shift={(0.7,0)}]
[name=meas]
dmeter q[1];
\end{scope}

\draw [-Stealth] (meas.north) |- ($(var.north) + (-4em,0.6em)$) -- ($(var.north) + (-4em,0)$) ;
\draw [-Stealth] (xj.south) -- (enc.north);
\end{yquant}
\end{tikzpicture}\end{small}
    \caption{Variational quantum circuit. The left part encodes input data, while the right part holds
    operators with trainable parameters. We abbreviate \(\xi\coloneqq \asin x\), and set \(\R{\sigma}{\theta}{k} \coloneqq R_{\sigma}(\theta_{k})\).
    The quantum register is set to $\ket{q} = \ket{0000}$}%
    \label{fig:variational_circ}
\end{figure}
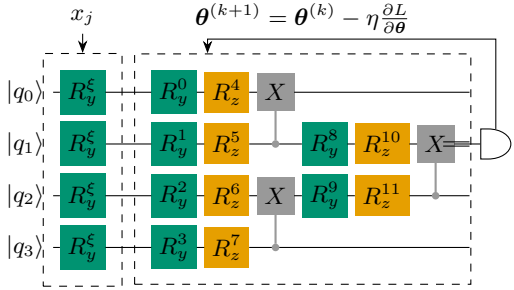

The previous examples are based on noise (inspired by the characteristics of real hardware),
which combines effects of different physical processes, as we have outlined above. We now
evaluate individual contributions of elementary noise types that  may provide guidance in
designing future quantum systems. We apply these to a shallow variational quantum circuit,
which is shown in~\autoref{fig:variational_circ}. Parameters are trained using the procedure
described in Sec.~\ref{sec:var_explanation}.\footnote{Note that the choice of
hyper-parameters (\eg, step size), as well as the classical optimiser influence algorithmic 
performance. While different hyper-parameters might be beneficial for different noise levels or
noise methods, we did not consider such an optimisation in the scope of this paper.}

\begin{figure*}[htbp]
    \includegraphics{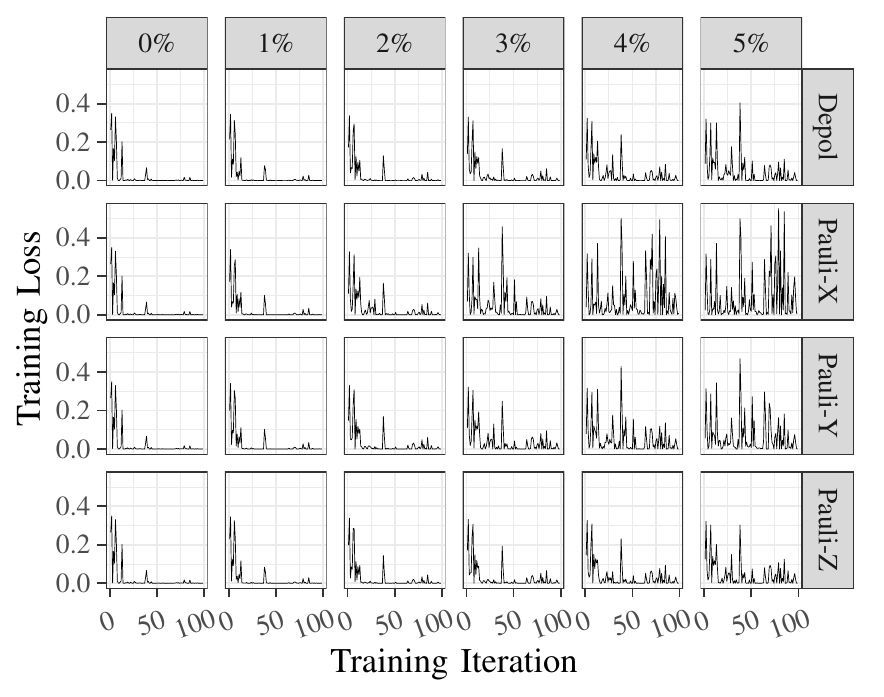}\includegraphics{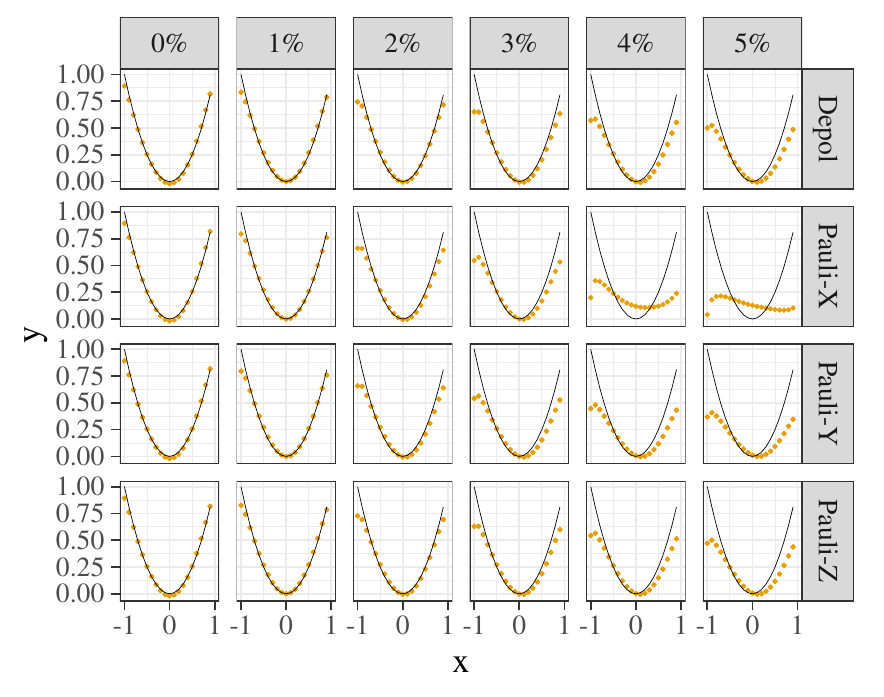}
    \caption{Approximation of $f(x) = x^2$ using the
    variational quantum circuit shown in Fig.~\autoref{fig:variational_circ}.
    (lhs) Comparison noise (applied in training and inference) variants effects on training loss.
    (rhs) Predictions (orange dots) versus target function
    (black line).
    }\label{fig:var_cmp_noise}
\end{figure*}

The training loss for Pauli-X, Y, and Z, as well as a depolarising channel, is shown in 
\autoref{fig:var_cmp_noise} for different levels of noise over 100 training iterations. The 
loss converges to zero
once parameters have been satisfactorily learned; the simulations show that increasing
noise strength impedes this process differently depending on the noise type. While
bit flips (Pauli-X) are particularly obstructive, platform
like trapped ion systems are particularly resilient against this type of noise~\cite{ebadi:2021,schymik:2020},
which might be a relevant criterion for choosing an underlying quantum platform for a
software system.

The right hand side of~\autoref{fig:var_cmp_noise} illustrates how predictions obtained
from the trained models degrade with increasing amounts of noise, providing tangible means 
of judging the effects of different types and strengths of noise. Both influence
result quality in different ways, and software engineers will need to decide (by choice of hardware) which 
variants are best tolerable for particular use-cases considering domain knowledge
and requirements.

\subsection{Connectivity and Gate-Sets}\label{sec:connectivity_gates}
Finally, let us elaborate on effects of connectivity structure and gate sets under the
influence of noise. For given architectures, the combination is fixed, but co-designing
systems that optimise either for specific algorithms is seen as one possible path
towards quantum advantage~\cite{wintersperger:22:codes,brown:2016,Li:2021,Feld:2018}. Software
engineers should be aware of possible future design opportunities.

\begin{figure}[htbp]
    \vspace*{-0.5em}\includegraphics{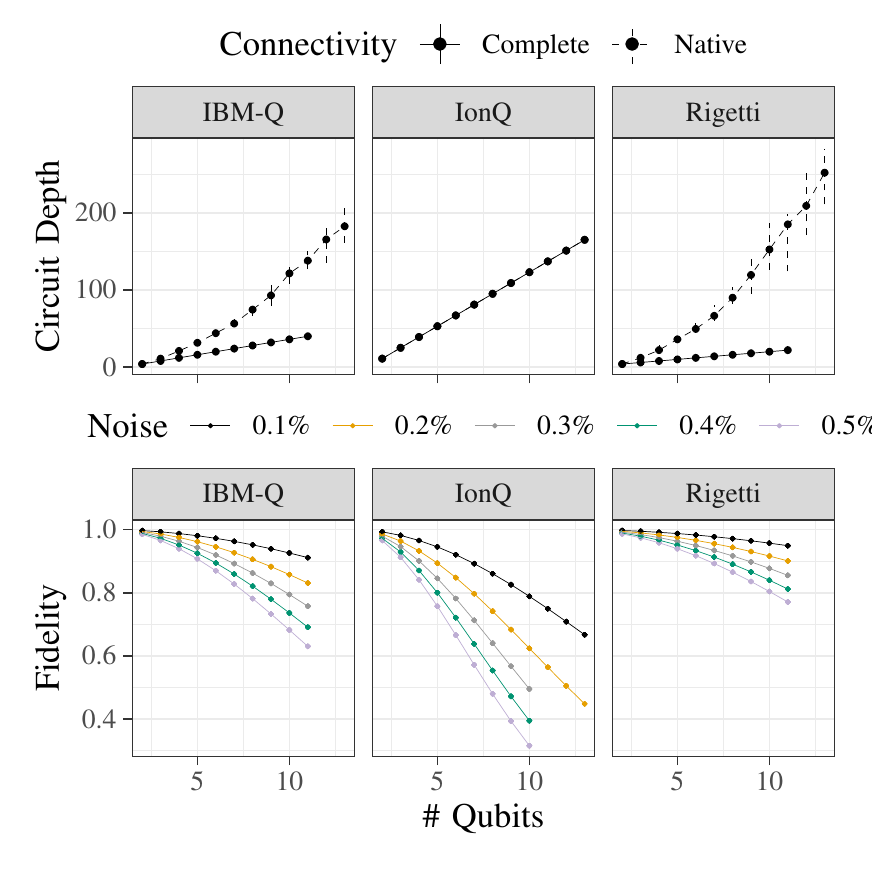}\vspace*{-1em}
    \caption{Impact of vendor-specific gate sets for QFT on (a) fully connected qubits versus
    vendor topologies (top), and (b) on result fidelity for  varying levels
    of depolarising noise and full connectivity (bottom). Circuit depths vary because of the stochastic transpilation.}\label{fig:gatesets}\vspace*{-1em}
\end{figure}

The top part of \autoref{fig:gatesets} fixes
the gate sets for each vendor, but shows circuit depth scaling for both, the native
connectivity structure and a fully connected
architecture (hypothetical for \ibmq and \rigetti, standard for \ionq). Owing to the need 
to insert swap
gates for \ibmq and \rigetti, depth grows super-linearly with native connectivity,
but increases linearly with a full mesh,
substantially reducing circuit depth.\footnote{Manufacturing a fully meshed
connectivity graph is extremely challenging for semiconductor-based
approaches. However, as Refs.~\cite{wintersperger:22:codes,safi:23:codesign} show,
even small additions to existing connectivity structures can result in major
improvements in circuit depth.}

For \ionq, circuit depth increases quicker than for the other architectures. This can
be attributed (based on manual inspection of the generated circuits) to weaknesses of the
circuit translator that maps logical to physical circuits,%
\footnote{Native \ionq compilers might improve results, yet do not satisfy our goal of providing open and reproducible means
of obtaining simulation results.} 
which stresses the
importance for software engineers to place greater emphasis on low-level details
like compiler performance that is only of marginal interest for many classical SE tasks.

The bottom part of~\autoref{fig:gatesets} illustrates some additional effects:
A fully coupled connectivity structure combined with fixed per-gate error rates for all vendors
isolates the effects of vendor specific base gates, and in particular, of compiling to them.
Here, the performance of \ionq base gates is mostly due to our sub-optimal transpiler,
as the resulting larger circuit depth gives more opportunity to \enquote{pick up noise}.
Software engineers must be aware of a possibly complex interplay of factors when
evaluating algorithmic quantum performance.

\section{Implications for Software Engineering}\label{sec:implications}
We have illustrated how noise and imperfections 
impact NISQ performance, and that a certain amount of knowledge of the underlying
mechanisms is necessary for proper interpretation. From the (quantum) software
engineering point of view, imperfections influence if and how non-functional
requirements can be satisfied.
In particular, they affect scalability, performance, testability, and cost.
The relation to the first two qualities has already been intensively
discussed above.

Testing outcomes of quantum algorithms needs to deal with two aspects of
uncertainty: Measurements leading
to stochastic outcome distributions, and imperfections in gates and components. While probabilistic behaviour is well established in classical computing~\cite{Mitzenmacher:2005}, physical imperfections
have found little consideration in SW testing to the best of our 
knowledge.
To appropriately design noise-aware
tests and judge test results, software engineers need to be able to understand quantum noise
at a sufficient level of detail.

The impact of noise and imperfections on quantum software could
be ignored if perfect, large-scale QPUs with error correction were available.
No physical reasons prevent designing the required systems, but many engineering 
challenges make specifying concrete roadmaps towards this goal challenging.
Even if such systems can eventually be built, practical industrial applications
will not just be judged by performance considerations, but by their overall 
cost-benefit trade-off, which is among core concern of any engineering discipline.
Consequently, since imperfect error correction and error mitigation schemes~\cite{Endo:2018}
will likely result in less costly machines, it seems reasonable
to assume that such machines will co-exist with perfect quantum computers, given 
they can solve certain tasks advantageously over classical computers. For instance, Liu~\etal~\cite{Liu:2021} prove exponential speedups for certain
types of quantum machine learning on fault-tolerant machines, which are believed to be extensible to NISQ machines using error-mitigation
(Hubregtsen~\etal~\cite{Hubregtsen:2022} study training embedding kernels on NISQ machines).

Some properties of quantum states and circuits require explicit consideration 
in designing new and extending existing test methodology: Not just the stochastic nature of
quantum measurements, but also the impact of imperfections makes defining desirable test results 
hard, as it is necessary to distinguish these measurement-induced variations
from variations induced by noise and imperfections. Guidelines that eliminate the need
for individual software engineers to be aware of statistical peculiarities could be established.

\section{Conclusion}\label{sec:discussion}
Using a reproducible and extensible empirical simulation approach, we illustrated
how noise and imperfections affect the the properties of quantum algorithms on
existing and hypothetical NISQ hardware. A solid understanding of
such effects is useful not only for researchers and engineers working on hardware
implementations, but also for software engineers. Yet, it seems unreasonable to equip
every software engineer or SWE researcher with detailed physical knowledge on noise.
We instead provide a suitably detailed introduction to the topic, accompanied by an easy-to-use replication package that allows software engineers
to explore the influence of noise with little effort. 
We deem this a crucial aspect in the endeavour of realising future quantum
applications of practical benefits.
\section*{Acknowledgements}
This work is supported by the German Federal Ministry of Education and Research within the funding program \emph{Quantentechnologien -- von den Grundlagen zum Markt}, contract number 13NI6092.

\IEEEtriggeratref{61}
\clearpage\bibliography{sources}
\bibliographystyle{IEEEtran}
\end{document}